\documentclass{article}
\usepackage{amsmath,amssymb,bm}
\usepackage{dsfont}
\usepackage{epsfig}
\usepackage{graphicx}
\usepackage{slashed}
\usepackage{color}
\usepackage{subfigure}

\newcommand{\be}{\begin{equation}}
\newcommand{\ee}{\end{equation}}

\newcommand{\beq}{\begin{equation}}
\newcommand{\eeq}{\end{equation}}
\newcommand{\bea}{\begin{eqnarray}}
\newcommand{\eea}{\end{eqnarray}}
\newcommand{\ba}{\begin{align}}
\newcommand{\ea}{\end{align}}
\newcommand{\bfig}{\begin{figure}}
\newcommand{\efig}{\end{figure}}

\newcommand{\as}{\alpha_s}

\newcommand{\nn}{\nonumber}

\begin{document}

\begin{center}

~\vspace{0.2cm}

{\bf DETERMINATION OF THE STRONG COUPLING FROM HADRONIC TAU DECAYS USING  RENORMALIZATION GROUP SUMMED \\PERTURBATION THEORY}\footnote{Contribution to the proceedings of the workshop "Determination of the Fundamental 
Parameters of QCD", Nanyang Technological University, Singapore, 18-22 March 2013, to be published in Mod. Phys. Lett. A.}
\\

\vspace{0.8cm}
GAUHAR ABBAS\footnote{Speaker}\\
\vspace{0.2cm}

The Institute of Mathematical Sciences, C.I.T.Campus,
Taramani, Chennai 600 113, India\\

\vspace{0.8cm}
 B. ANANTHANARAYAN\\
\vspace{0.2cm}

Centre for High Energy Physics,
Indian Institute of Science, Bangalore 560 012, India\\

\vspace{0.8cm}
 I. CAPRINI\\
\vspace{0.2cm}

Horia Hulubei National Institute for Physics and Nuclear Engineering,\\
P.O.B. MG-6, 077125 Bucharest-Magurele, Romania

\end{center}
\vspace{0.2cm}

\begin{abstract}
We determine the strong coupling constant $\alpha_s$ from the $\tau$ hadronic
 width using a renormalization group summed (RGS) expansion of
 the QCD Adler function. The main theoretical uncertainty in the extraction
 of $\alpha_s$ is due to the manner in which renormalization
 group invariance is implemented, and the as yet uncalculated higher order
 terms in the QCD perturbative series. We show
 that new expansion exhibits good renormalization group improvement and the
 behaviour of the series is similar to that of the standard
 CIPT expansion. 
The value of the strong coupling in 
${\overline{\rm MS}}$ scheme obtained with the RGS expansion is $ \alpha_s(M_\tau^2)= 0.338 \pm 0.010$.  
The convergence properties of the new 
 expansion can be improved by Borel transformation and 
 analytic continuation in the Borel plane.  This is discussed elsewhere in these proceedings.

\end{abstract}

\section{Introduction}	

The inclusive hadronic decay width of the $\tau$ lepton 
provides a  clean source for the  determination of  $\alpha_s$ at
low energies \cite{RPP,Pich:2013sqa,Jamin:2013hll,Altarelli:2013bpa}.  
 The perturbative QCD contribution is known up to $O(\alpha_s^4)$ \cite{BCK08} and the 
nonperturbative corrections are found to be 
small \cite{BrNaPi,Davier2006,Davier2008,BeJa,Pich2010,Pich_Muenchen}.
The main uncertainties arise from the treatment of higher-order 
corrections and  improvement of the perturbative series through
renormalization group (RG) methods.  The 
leading methods for the treatment of the perturbative series 
are fixed-order perturbation theory (FOPT) and 
contour-improved perturbation theory (CIPT)~\cite{Pivovarov:1991rh,dLP1}.  
The predictions from these methods are not in
agreement and the discrepancy between them is one of the main sources of ambiguity 
in the extraction of 
 $\alpha_s$ \cite{BeJa,Pich_Muenchen,Pich_Manchester,Bethke,Beneke_Muenchen}.  

The above theoretical discrepancy has been the motivation for proposing an alternative approach.  
We consider the method developed in \cite{Ahmady1,Ahmady2}, 
using a procedure originally 
suggested in \cite{MaMi1,Maxwell,MaMi2} 
which we refer to as  renormalization-group summation (RGS).  
This  is a framework involving leading logarithms summation, 
in which terms in powers of the coupling constant and logarithms are 
regrouped, so that for a given order, the new expansion includes 
every term in the perturbative series that can be calculated 
using the RG invariance.  
The results, which are summarized in this talk, are similar to those in CIPT and predicts $\alpha_s$ close to the CIPT prediction \cite{Abbas:2012py}.
Note that this method was used for the study of the inclusive decays 
of the b-quark and the hadronic cross section in 
$e^+ e^-$ annihilation \cite{Ahmady1,Ahmady2}. 
Our work demonstrates that the method can also be applied with  success to 
the determination of $\alpha_s$ from $\tau$ hadronic decays.

It must be noted that the QCD perturbative  corrections are 
also sensitive to as yet uncalculated higher order terms in the series.  
The coefficients of the perturbative series grow as $n!$  and the series have zero radius of convergence\cite{Muell,Muell1,Broa,Bene}.  
Consequently one can study
the Borel transform of the QCD Adler function which has ultraviolet (UV) 
and infrared (IR) renormalon singularities in the Borel plane.  
The divergent behaviour can be considerably tamed by using 
techniques of series acceleration based on conformal mappings 
and `singularity softening' \cite{CaFi1998,CaFi2000,CaFi2001,CaFi2009,CaFi2011,CaFi_Manchester}. 
The method is not applicable to the perturbative series 
in powers of $\alpha_s$ since the expanded correlators are 
singular at $\alpha_s=0$, but can be applied in the Borel plane. 
The large order behaviour of the RGS expansion can be 
improved using this method \cite{Abbas:2012fi}.
This provides  a new class of expansions, 
which simultaneously implement the RG invariance and the 
large-order summation by 
the analytic continuation in the Borel plane. The details may be found elsewhere in these proceedings \cite{capriniqcd13}.

\section{The QCD Adler function}	
Our treatment begins with the QCD Adler function, which
enters the expression for the total inclusive hadronic width
of the $\tau$ lepton.
The total inclusive  hadronic width of the  $\tau$ 
provides an accurate calculation of the ratio 
\begin{equation}
\label{Rtauex}
R_{\tau} \,\equiv\, \frac{\Gamma[\tau^- \to {\rm hadrons} \, \nu_\tau ]}
{\Gamma[\tau^- \to e^- \overline \nu_e \nu_\tau ]} .
\end{equation}
The Cabibbo-allowed combination $R_{\tau,V+A}$ which proceeds through a vector and an axial vector current 
can be written as 
\begin{equation}
\label{eq:delta}
   R_{\tau, V+A}  =
      S_{\rm EW} |V_{ud}|^2\bigg(1 + \delta^{(0)} +  \delta_{\rm EW} + \delta^{(2,m_q)}_{ud,V/A}
+\sum_{D=4,6,\dots} \delta_{ud,V/A}^{(D)}\bigg)\,,
\end{equation}
where $S_{\rm EW}$  and
$\delta_{\rm EW}$  are electroweak corrections, $\delta^{(0)}$ is the dominant universal perturbative QCD correction, and $\delta_{ud}^{(D)}$ denote quark mass
and higher $D$-dimensional operator corrections (condensate contributions) arising in
the operator product expansion (OPE).  
Our main interest is in the perturbative correction $\delta^{(0)}$ which can be written as
\begin{equation}
\label{del01}
\delta^{{(0)}}=\frac{1}{2 \pi i}\!\! \oint\limits_{|s|=M_\tau^2}\!\! \frac{d s}{s}
\left(1- \frac{s}{M_\tau^2}\right)^3 \left(1+\frac{s}{M_\tau^2}\right) \hat {D}_{\rm pert}(s),
\end{equation}
 where 
 $\hat D_{\rm pert}(s)\equiv D^{(1+0)}(s)-1$ is the perturbative part of the reduced function Adler function.
The perturbative expansion of $\widehat{D}(s)$  in the  ``fixed-order perturbation theory" at $\mu^2=M_\tau^2$  reads \cite{BeJa}
\begin{equation}
\label{DsFo}
 \widehat{D}_{\rm FOPT}(a_s, L) = \sum\limits_{n=1}^\infty a_s^n
\sum\limits_{k=1}^{n} k\, c_{n,k}\,L^{k-1} \,.
\end{equation}
where
\beq\label{aL}
 a_s\equiv \as(\mu^2)/\pi,\quad\quad  L\equiv\,\ln (-s/\mu^2).
\eeq
The  coefficients $c_{n,1}$ in the ${\overline{\rm MS}}$ scheme for $n_f=3$ flavours are  (cf. \cite{BCK08} and references therein):
\beq\label{cn1}
c_{1,1}=1,\, c_{2,1}=1.640,\, c_{3,1}=6.371,\, c_{4,1}=49.076.
\eeq
Several estimates for the next coefficient $c_{5,1}$ are available \cite{Davier2008,BeJa,Pich_Muenchen,Beneke_Muenchen}.
The remaining coefficients $c_{n,k}$ for $k>1$  are determined from renormalization-group invariance.  The function   $\widehat{D}_{\rm pert}$ is scale independent 
and therefore satisfies  the RG equation 
\begin{equation}\label{eq:rgi}
\mu^2 \frac{\mathrm{d}}{\mathrm{d}\mu^2} \left[  \widehat{D}_{\rm pert}(s) \right] =
0,
\end{equation}
which takes the form
\begin{equation}\label{pde2}
 \beta(a_s)\frac{\partial \widehat{D}_{\rm pert} (s)}{\partial a_s}-\frac{\partial \widehat{D}_{\rm pert} (s)}{\partial \ln (-s/\mu^2) }  = 0,
\end{equation}
where
\begin{equation}\label{beta}
\beta(a_s) \equiv \mu^2 \frac{\mathrm{d}a_s(\mu^2)}{\mathrm{d}\mu^2} = -(a_s(\mu^2))^2 \sum_{k=0}^\infty \beta_k (a_s(\mu^2)) ^k
\end{equation}
is the  $\beta$ function. The known $\beta_j$  coefficients are \cite{LaRi,LaRi2,Czakon}:
\beq\label{betaj}
 \beta_0=9/4,\,\, \beta_1=4,\,\, \beta_2= 10.0599,\,\,\beta_3=47.228.
\eeq

The series (\ref{DsFo}) is badly behaved especially near the time-like axis due to the large imaginary part of  
the logarithm  $\ln (-s/ M_\tau^2)$  along the circle $|s|= M_\tau^2$ \cite{Pivovarov:1991rh,dLP1}.
The choice $\mu^2=-s$ provides the CIPT expansion \cite{Pivovarov:1991rh,dLP1}
\beq\label{DsCI}
 \widehat{D}_{\rm CIPT}(s)= \sum\limits_{n=1}^\infty  c_{n,1} (a_s(-s))^n
\,,
\eeq
where the running coupling  $a_s(-s)$  is determined by solving the renormalization-group equation (\ref{beta}) numerically in an iterative way along the circle, 
starting with the input value $a_s(M_\tau^2)$ at $s=-M_\tau^2 $.  This expansion avoids the appearance of large logarithms along the circle $|s|= M_\tau^2$. 
\section{Renormalization-Group Summation}
The  expansion of the Adler function (\ref{DsFo})  can be written in the RGS form \cite{Abbas:2012py,Abbas:2012fi}
\begin{equation}
\label{dseries}
 \widehat{D}_{\rm FOPT}(a_s, L) =
\sum_{n=1}^\infty a_s^n D_n (a_sL),
\end{equation}
 where the functions $D_n(a_s L )$, depending on a single variable $u=a_sL$, are defined as 
\begin{equation}\label{Dn_def}
D_n (u) \equiv \sum_{k=n}^\infty (k-n+1)c_{k, k-n+1} u^{k-n}.
\end{equation}

The function $D_1$ sums all the leading logarithms, the second function $D_2$ sums the next-to-leading logarithms, and so on.  These function 
can be obtained in a closed analytical form using RG invariance through the equation (\ref{pde2}) and the expansion (\ref{dseries}).  This gives the following RGE equation
\begin{align}
0 & = -\sum_{n=1}^\infty
\sum_{k=2}^n k (k-1) c_{n,k} \, a_s^n L^{k-2}\nonumber \\
& - \left(\beta_0 a_s^2 + \beta_1 a_s^3 + \beta_2 a_s^4  + 
\ldots +\beta_l a_s^{l+2} +\ldots \right)  \times\sum_{n=1}^\infty
\sum_{k=1}^n n k c_{n,k}  a_s^{n-1} L^{k-1}.
\end{align}
By extracting the aggregate coefficient of $a_s^n L^{n-p}$  one obtains the recursion formula
$(n \geq p)$
\begin{equation}\label{recursion}
0 = (n-p+2) c_{n, n-p+2}+ \sum_{\ell = 0} ^{p-2}  (n - \ell - 1) \beta_\ell c_{n - \ell - 1, n - p+1}.
\end{equation}
Multiplying both sides of (\ref{recursion})  by $(n-p+1) u^{n-p}$  and summing from $n=p$ to
$\infty$, we obtain a set of first-order linear differential equation for the functions 
defined in (\ref{Dn_def}) for  $n\ge 1$: 

\beq\label{Dk_de}
\frac{\mathrm{d}D_n}{\mathrm{d}u} +  \sum_{\ell = 0}^{n-1} \beta_\ell \left( u \frac{\mathrm{d}}{\mathrm{d}u} + n - \ell \right) D_{n - \ell}=0,
\eeq
with the initial conditions $D_n (0) = c_{n,1}$ which follow from (\ref{Dn_def}). 
The solution of the system (\ref{Dk_de}) can be found iteratively in an analytical form.  The solutions $D_n (u)$ depend on the coupling constant and 
logarithms.  The expressions of $D_{n}(u)$ for $n \leq 4 $, written in terms of the coefficients $c_{k,1}$ with $k\leq n$ and $\beta_j$ with $0\le j\le n-1$, are:
\bea\label{D12}
D_1 (u) && = \frac {c_{1, 1}} {w},\quad\quad\quad w=1 + \beta_0 u,\nonumber \\
D_2 (u) && =   \frac {c_{2, 1}} {w^2} - \frac {\beta_{1} c_{1,1} \ln w} {\beta_{0} w^2},
\eea
\begin{align}
\label{D3}
D_3(u) & =  \frac {(\beta_ {1}^2 - \beta_ {0}\beta_ {2}) c_ {1, 
    1} } {\beta_ {0}^2  w^{2} } +  \left [\frac {(-\beta_{1}^2 
 + \beta_{0} \beta_{2}) c_{1, 
     1}} {\beta_{0}^2}  + 
 c_{3, 1} \right] w^{-3}
\\
\nonumber & 
 + \left [-\frac {\beta_{1} (\beta_{1} c_{1, 1} + 
      2\beta_{0} c_{2, 
        1}) \ln w} {\beta_{0}^2} 
+ \frac {\beta_{1}^2 c_{1, 1} \ln^2 w} {\beta_{0}^2}\right] w^{-3}. 
\end{align}

\bea
&&D_4 (u) =  -\frac {(\beta_{1}^3 - 
2\beta_{0} \beta_{1} \beta_{2} + \beta_{0}^2\beta_{3}) c_{1, 1}} {2\beta_ {0}^3} w^{-2}
-\left [\frac {\beta_{1} (-\beta_{1}^2 + \beta_{0} \beta_{2}) c_{1,1}} {\beta_{0}^3} + 
\frac {2 (-\beta_{1}^2 + \beta_{0} \beta_{2}) c_{2,1}} {\beta_{0}^2} \right] w^{-3} \nn
\nonumber \\
&& + \frac {2\beta_{1} (-\beta_{1}^2 + \beta_{0} \beta_{2}) c_{1, 1} 
\ln w} {\beta_{0}^3}w^{-3}
+  \left[\frac {(-\beta_{1}^3 + \beta_{0}^2\beta_{3}) c_{1,1}} 
{2\beta_{0}^3} + \frac {2 (-\beta_{1}^2 + \beta_{0} \
\beta_{2}) c_ {2,1}} {\beta_{0}^2} + c_{4,1}\right]w^{-4} 
\nonumber \\
&&-  \frac {\beta_{1} (-2 \beta_{1}^2 c_{1, 1} + 
3 \beta_{0} \beta_{2} c_{1, 1} + 
2 \beta_{0} \beta_{1} c_{2, 1} + 
3 \beta_{0}^2 c_{3, 1}) \ln w} {\beta_{0}^3}w^{-4}
\nonumber \\
&&
+  \frac {\beta_{1}^2 (5\beta_{1} c_{1, 1} + 
6\beta_{0} c_{2,1}) \ln^2 w} {2\beta_{0}^3}w^{-4}
- \frac {\beta_{1}^3 c_{1,1} \ln^3 w} {\beta_{0}^3} w^{-4}.\nn
\eea
The higher order solutions can be found in 
Ref. \cite{Abbas:2012py}.  These expressions are used
for computing the perturbative contribution to the hadronic width of the $\tau$ lepton and
the subsequent extraction of $\alpha_s(M_\tau^2)$.
\section{The properties of the RGS expansion}
\begin{figure}[thb]
 	\begin{center}\vspace{0.0cm}
 	 \includegraphics[width = 7.0cm]{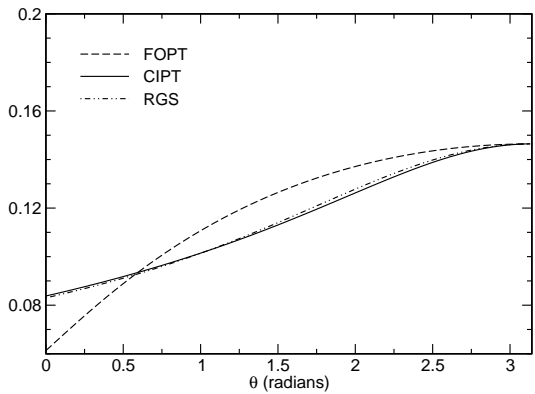}
	\caption{ Modulus of the Adler function expansions (\ref{DsFo}),  (\ref{DsCI}) and  (\ref{dseries}), summed up to the order $N=5$, along the circle $s=M_\tau^2 \exp(i\theta)$. }
	\label{fig1}
 	\end{center}\vspace{0.0cm}
\end{figure}

We now discuss the properties of the RGS expansion in the complex momentum plane, along the circle $s= M_{\tau}^2 \exp (i \theta)$. In 
Fig. \ref{fig1}, we show the behaviour of the modulus  of the Adler function along the circle given by the first $N=5$ terms in the expansions (\ref{DsFo}), (\ref{DsCI}) and (\ref{dseries}), respectively. In this calculation and below we used the standard value $\as(M_{\tau}^2)=0.34$, adopted also in previous studies \cite{BeJa,CaFi2011}.  One may see that the behaviour of the new RGS expansion 
is similar to that of the CIPT  expansion. 

In Table \ref{tab1}, we show the values of the quantity $\delta^{(0)}$ defined in (\ref{del01}), calculated with FOPT, CIPT and RGSPT, as a function of the perturbative order up to which the series was summed.  We observe that CIPT 
shows a faster convergence compared to FOPT.  To order $N=4$, the difference between FOPT and CIPT is $0.0215$, which is  the main source of the 
theoretical  uncertainty in the extraction of $\alpha_s$ from the hadronic $\tau$ decay rate. 
The new RGS expansion, as remarked earlier, shows the convergence which is similar to the CIPT expansion.  We note that 
for $N=4$, the difference between the results of the RGS and the 
standard  FOPT is $0.01754$, and the difference from  the RGS and CIPT is $0.0039$, which confirms that the new expansion gives results close to those of the CIPT. 
For $N=5$, using the estimate $c_{5,1}=283$ from \cite{BeJa}, we find that the RGS differs from FOPT by $0.0232$, and from  CIPT by $0.0035$.

\begin{table}[h]
\begin{center}
\caption{Predictions of $\delta^{(0)}$ by the standard FOPT, CIPT and  the RGS,
 for various truncation orders $N$.}
{\begin{tabular}{lccc}\hline\hline 
$ $ & $\delta^{(0)}_{\rm FOPT}$ & $\delta^{(0)}_{\rm CIPT}$   & $\delta^{(0)}_{\rm RGS}$ \nonumber \\ \hline 
$ N=1 $ &~~ 0.1082~~ & ~~0.1479~~ & ~~ 0.1455 ~~ \nonumber \\ 
$ N=2 $ & 0.1691  & 0.1776 &  0.1797 \nonumber \\ 
$ N=3 $ & 0.2025 & 0.1898 &  0.1931 \nonumber \\ 
$ N=4 $ & 0.2199  & 0.1984   & 0.2024\nonumber \\ 
$ N=5 $ & 0.2287   & 0.2022     &  0.2056 \nonumber \\  \hline 
\end{tabular}
\label{tab1} }
\end{center}
\end{table}

It would be of interest to study the behaviour of RGS expansion at higher orders which are not shown in the Table \ref{tab1}.  This is the subject of the next section, in a  
model for higher order coefficients of the Adler function.

\section{Higher order behaviour of the RGS expansion}
\begin{figure}[thb]
 	\begin{center}\vspace{0.5cm}
 	 \includegraphics[width = 7.3cm]{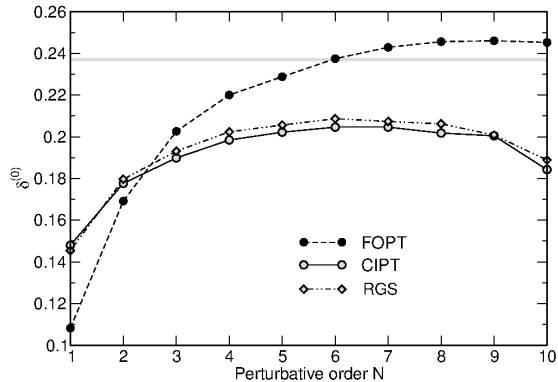}
	\caption{Dependence of $\delta^{(0)}$ in FOPT, CIPT and RGS  on the truncation order $N$ in the Beneke and Jamin model \cite{BeJa}. The gray band is the exact value. }
	\label{fig6}
 	\end{center}\vspace{0.0cm}
\end{figure}

As discussed earlier, the extraction of $\alpha_s$ from the hadronic $\tau$ decays width is also sensitive to the large order behaviour of the QCD 
perturbative series.  It is of interest to check if the low order behaviour  of the new RGS expansion persists at higher orders.  This investigation was carried out in 
\cite{Abbas:2012py,Abbas:2012fi} in a model proposed in \cite{BeJa}.
In this model, the RGS expansion of the QCD Adler function 
has a behaviour which is similar to that of 
CIPT and eventually exhibits  big  oscillations, 
showing the divergent character of the QCD perturbative series 
at higher orders. 
In Fig. \ref{fig6}, we show the behaviour of FO, CI and RGS expansions in 
the so-called ``reference model" defined in \cite{BeJa,Beneke:2012vb}. The RGS  results are
close to those of CIPT at every order up to $N=10$.  
In this model,  FOPT expansion approaches better the `true value'. 

A method
for taming the divergent behaviour of the QCD perturbative expansions was proposed in
\cite{CaFi1998,CaFi2000,CaFi2001,CaFi2009,CaFi2011,CaFi_Manchester}, using the series acceleration by
the conformal mappings of the Borel plane and 
the implementation of the known nature of the leading singularities in this plane. 

In Ref. \cite{Abbas:2012fi}, we applied these techniques to
the RGS expansion and defined a novel, RGS non-power (RGSNP) expansion of the QCD Adler function, in which the powers of the coupling of the standard expansion are replaced by 
suitable functions that resemble the expanded amplitude as concerns their singularities and the expansions in powers of $\alpha_s$.  
The non-power expansions 
show remarkable convergence properties for the Adler function, which is the crucial input in the determination of $\alpha_s$ from hadronic $\tau$ decays. This is reviewed elsewhere 
in these proceedings \cite{capriniqcd13}.

\section{Determination of $\alpha_s (M_{\tau}^2)$ from RGS expansion }
In this section, we present the derivation of 
$\alpha_s(M_\tau^2)$ following Ref. \cite{Abbas:2012py}.  We use  as input the coefficients   $c_{n,1}$  calculated from Feynman diagrams, given in (\ref{cn1}), and the estimate  $c_{5,1}=283\pm 283$ \cite{BeJa,Beneke_Muenchen}.
We need also the  phenomenological value of the pure perturbative correction to the hadronic $\tau$ width, for which we adopt the recent estimate \cite{Beneke_Muenchen}, 
\begin{equation}\label{delph}
 \delta^{(0)}_{\rm phen}=0.2037\pm 0.0040_{exp}\pm 0.0037_{\rm PC},
\end{equation}
where the first error is experimental and the second shows the uncertainty due to the power corrections. 
This value has been used in several recent determinations \cite{CaFi2009,Beneke_Muenchen,CaFi2011}.
 With 
this input we obtain 
\bea\label{alpha}
\alpha_s(M_\tau^2)&=&0.3378  \pm  0.0046_{\rm exp}  \pm 0.0042_{\rm PC} ~^{+0.0062}_{- 0.0072}(c_{5,1})\nn
\nonumber \\&&^{+ 0.0005}_{-0.0004 }{(\rm scale)} \pm ^{+ 0.000085}_{-0.000082}{(\rm \beta_4)}.
\eea
In the above, the first two errors are due to the corresponding uncertainties of $\delta^{(0)}_{\rm phen}$ given in (\ref{delph}), while the third is due
to the uncertainty of the 
coefficient  $c_{5,1}$ with the conservative range adopted above, the fourth is due to scale variation, and the last one is due to the effect of 
the truncation of the $\beta$-function expansion.  The details may be found in \cite{Abbas:2012py}.
We observe that $\alpha_s(M_\tau^2)$ is not sensitive to
 the  variation of the scale.  The largest error comes from the 
uncertainty  of the five loop coefficient  $c_{5,1}$.  This was also observed in the standard CIPT analysis \cite{Pich_Muenchen,Pich_Manchester} and in the analysis based on
the CI expansions improved by the conformal mappings of the Borel plane \cite{CaFi2009,CaFi2011}. 

Combining in quadrature the errors given in (\ref{alpha}), the  prediction based on RGS expansion reads  \cite{Abbas:2012py} 
\begin{equation}\label{aver1}
\alpha_s(M_\tau^2)= 0.338 \pm 0.010.
\end{equation}
We mention that for the same input (\ref{delph}) the standard FOPT and CIPT give, respectively, 
\bea\label{FOCI}
\alpha_{s}(M_{\tau}^2)&=& 0.320^{+0.012}_{-0.007}, \quad\quad\quad{\rm FOPT}, \nonumber \\ \alpha_{s}(M_{\tau}^2)&=& 0.342 \pm  0.012, \quad\quad{\rm CIPT}.
\eea
For comparison we mention also the value  $\alpha_{s}(M_{\tau}^2)= 0.320^{+0.019}_{-0.014}$, obtained recently in Ref. \cite{CaFi2011} with the same input (\ref{delph})  and the 
CI non-power  (CINP) expansion  and the value $\alpha_s(M_\tau^2)= 0.319~^{+ 0.015}_{-0.012 }$ determined by 
the RGS non-power (RGSNP) expansion  in the Ref. \cite{Abbas:2012fi} based on the analytic continuation in the Borel plane.\footnote{The question of the uncertainties due to the
 nonperturbative corrections has been recently addressed in  detail in
the Ref. \cite{Boito:2012cr}, where an error larger than that quoted for power corrections 
in the Eq. (\ref{delph}) has been obtained.  Using this more conservative input will slightly increase the error in the predictions above. For instance, as
discussed in \cite{Abbas:2012fi}, for the  RGSNP prediction  the upper and lower
errors change to $0.017$  and $0.015$ respectively.}
  After evolving  to the scale of $M_Z$, 
the RGSNP prediction of  $\alpha_s$ reads $ \alpha_s(M_Z^2)= 0.1184~^{+0.0018}_{-0.0015}$.   
The special features of this latter expansion will be discussed elsewhere in these proceedings \cite{capriniqcd13}.

\section{Conclusion}

In this talk, we have presented our recent results on the determination of the strong coupling constant $\alpha_s$ from the $\tau$ hadronic width,  based on the formalism which we denote as RGS expansion of the Adler function.  
Due to the present discrepancy in the determination of the $\alpha_s$ from hadronic $\tau$ decays, any alternative approach 
besides FOPT and CIPT, 
must be pursued.  It must, however, be noted that the RGS framework is an important
 framework, which was simply not explored in detail earlier in the context of
the hadronic decay of the $\tau$ lepton.  Our work fills this gap.
The method discussed in this talk exploits RG invariance in a complete
way, providing analytical closed form solutions to a definite order.  The truncated summation of the perturbative series 
differ among each other by terms of the order $\alpha_s^{N+1}$.  
This difference turns out to be quite important for the low scale  relevant in $\tau$ decays. 

We have discussed in detail the properties of the RGS expansion,
including its properties in the complex energy plane and 
concluded that these properties are similar to those of CIPT expansion.  
We also provide the 
value of the strong coupling $\alpha_s$ from this expansion, 
which is closer to CIPT prediction.
We conclude that the  the summation of leading logarithms provides a systematic expansion with good convergence properties in the complex plane, including the critical 
region near the time-like region.  

The determination of  $\alpha_s$ is also ambiguous due to the effect of the as yet uncalculated higher order terms in the perturbative expansion of the hadronic width.  This ambiguity 
is amplified by the fact that 
the perturbative series is divergent, the coefficients displaying a factorial increase.  In QCD, due to the presence of the UV and  IR renormalon singularities situated on the real 
axis in the Borel plane, the Borel-Laplace integral   giving the expanded correlators in terms of their Borel 
transform  requires a prescription. Using the Principal Value prescription and the technique of series acceleration by conformal mappings and singularity softening in the Borel plane 
developed in \cite{CaFi1998} - \cite{CaFi_Manchester}, we have defined in \cite{Abbas:2012fi}  a new kind of expansion,  referred to as RGS non-power (RGSNP) expansion.   The divergent 
behaviour of the  standard perturbative series is 
considerably tamed in the non-power expansions, which show good convergence in the complex energy plane.   
More details may be found  in Refs. \cite{Abbas:2012fi,capriniqcd13}. 

\section*{Acknowledgments}
We thank Prof. K. K. Phua and the workshop organizers for the kind hospitality
at the Institute for Advanced Studies, Nanyang Technological University,
Singapore. IC acknowledges  support from the Ministry of National Education
under Contracts PN 09370102/2009 and Idei-PCE No 121/2011.


\begin{thebibliography}{0}

\bibitem{RPP} 
  J. Beringer {\it et al.}  (Particle Data Group),
  Phys.\ Rev.\ D{\bf 86}, 010001 (2012).
\bibitem{Pich:2013sqa} 
  A.~Pich, {\it Review of $\alpha_s$ determinations},
  arXiv:1303.2262.


\bibitem{Jamin:2013hll} 
  M.~Jamin,{\it Determination of $\alpha_s$ from tau decays},
  arXiv:1302.2425.

\bibitem{Altarelli:2013bpa} 
  G.~Altarelli,{\it The QCD Running Coupling and its Measurement},
  arXiv:1303.6065.


\bibitem{BCK08} P.A. Baikov, K.G. Chetyrkin and J.H. K{\"u}hn, 
 Phys. Rev. Lett. {\bf 101}, 012002  (2008), arXiv:0801.1821.
\bibitem{BrNaPi}
  E. Braaten, S. Narison and A. Pich,
  Nucl.\ Phys.\  B {\bf 373}, 581 (1992).



\bibitem{Davier2006}  M. Davier, A. H\"ocker and Z. Zhang,   Rev. Mod. Phys. {\bf 78}, 1043 (2006), hep-ph/0507078.


\bibitem{Davier2008} M. Davier, S. Descotes-Genon, A. H\"ocker, B. Malaescu and Z. Zhang,  
Eur. Phys. J. C{\bf 56}, 305  (2008), arXiv:0803.0979.

\bibitem{BeJa} M. Beneke and M. Jamin, JHEP {\bf 09}, 044 (2008), arXiv:0806.3156. 




\bibitem{Pich2010} A. Pich,  Acta Phys. Polon. Supp. 3,  165 (2010), arXiv:1001.0389. 
\bibitem{Pich_Muenchen}  A. Pich, {\it Tau decay determination of the QCD coupling}, in
 {\it Workshop on Precision Measurements 
of $\alpha_s$},  ed. S. Bethke {\it et al}, page 21, arXiv:1110.0016.


\bibitem{Pivovarov:1991rh} 
  A.A.Pivovarov,
  Z.\ Phys.\ C {\bf 53}, 461 (1992),
  [Sov.\ J.\ Nucl.\ Phys.\  {\bf 54}, 676 (1991)]
  [Yad.\ Fiz.\  {\bf 54}, 1114 (1991)];
  hep-ph/0302003.

\bibitem{dLP1}
  F. Le Diberder and A. Pich,
  Phys.\ Lett.\  B {\bf 286}, 147 (1992).



\bibitem{Pich_Manchester}  A. Pich, Nucl. Phys. B Proc. Suppl., 218, 89 (2011), 
arXiv:1101.2107.  

\bibitem{Bethke} S. Bethke,  Eur. Phys. J. C{\bf 64}, 689 (2009), arXiv:0908.1135.

\bibitem{Beneke_Muenchen} M. Beneke and M. Jamin, {\it Fixed-order analysis of the hadronic $\tau$ decay width}, in  
{\it Workshop on Precision Measurements 
of $\alpha_s$},  ed. S. Bethke {\it et al}, page 25, arXiv:1110.0016.


\bibitem{Ahmady1}
M.R. Ahmady, F.A. Chishtie,  V. Elias, A.H. Fariborz, N. Fattahi, D.G.C. McKeon, T.N. Sherry, T.G. Steele,
  Phys.\ Rev.\  D {\bf 66}, 014010 (2002),
  hep-ph/0203183.
\bibitem{Ahmady2}
M.R. Ahmady, F.A. Chishtie, V. Elias, A.H. Fariborz, D.G.C. McKeon, T.N. Sherry, A. Squires, T.G. Steele,
  Phys.\ Rev.\  D {\bf 67}, 034017 (2003),
  hep-ph/0208025.
\bibitem{MaMi1} C.J. Maxwell and A. Mirjalili,  Nucl.Phys. B{\bf 577}, 209 (2000),
hep-ph/0002204.

\bibitem{Maxwell} 
C.J. Maxwell, Nucl.Phys.Proc.Suppl. {\bf 86},  74 (2000).
\bibitem{MaMi2} C.J. Maxwell and A. Mirjalili, Nucl.Phys. B{\bf 611}, 423 (2001), 
hep-ph/0103164.



\bibitem{Abbas:2012py} 
  G.~Abbas, B.~Ananthanarayan and I.~Caprini,
  Phys.\ Rev.\ D {\bf 85}, 094018 (2012),
  arXiv:1202.2672.

\bibitem{Muell}  A.H. Mueller,
  Nucl.\ Phys.\ B {\bf 250}, 327 (1985).

\bibitem{Muell1} A.H. Mueller, in {\em QCD - Twenty Years Later}, Aachen 1992, 
edited by P. Zerwas and H. A. Kastrup (World Scientific, Singapore, 1992).

\bibitem {Broa} D. Broadhurst, Z.\ Phys.\ C {\bf 58}, 339 (1993).
\bibitem {Bene} M. Beneke, Nucl.\ Phys.\ B {\bf 405}, 424 (1993).

\bibitem{CaFi1998}
  I. Caprini and J. Fischer,
  Phys.\ Rev.\  D {\bf 60}, 054014 (1999),
  hep-ph/9811367.

\bibitem{CaFi2000}
  I. Caprini and J. Fischer,
  Phys.\ Rev.\  D {\bf 62}, 054007 (2000),
  hep-ph/0002016.

\bibitem{CaFi2001}
  I. Caprini and J. Fischer,
  Eur.\ Phys.\ J.\  C {\bf 24}, 127 (2002),
  hep-ph/0110344.

\bibitem{CaFi2009} I. Caprini and J. Fischer,   Eur. Phys. J. C{\bf 64}, 35  (2009), 
arXiv:0906.5211.
\bibitem{CaFi2011}
  I. Caprini and J. Fischer, 
Phys. Rev. D {\bf 84}, 054019 (2011),
arXiv:1106.5336. 

\bibitem{CaFi_Manchester} I. Caprini and J. Fischer, Nucl. Phys. B Proc. Suppl., 218, 
128 (2011), arXiv:1011.6480. 


\bibitem{Abbas:2012fi} 
  G.~Abbas, B.~Ananthanarayan, I.~Caprini and J.~Fischer,
  Phys.\ Rev.\ D {\bf 87}, 014008 (2013),
  arXiv:1211.4316.

\bibitem{capriniqcd13} 
   I. Caprini,  in this issue of Mod. Phys. Lett. A. 
 



\bibitem{LaRi} S.A. Larin, T. van Ritbergen and J.A.M. Vermaseren, Phys.
Lett. B{\bf 400}, 379 (1997), hep-ph/9701390.
\bibitem{LaRi2}
S.A. Larin, T. van Ritbergen and J.A.M. Vermaseren,
Phys. Lett. B{\bf 404}, 153 (1997),hep-ph/9702435.

\bibitem{Czakon}   M. Czakon,  Nucl.\ Phys. B{\bf 710}, 485 (2005),hep-ph/0411261.



\bibitem{Beneke:2012vb} 
  M.~Beneke, D.~Boito and M.~Jamin,
  JHEP {\bf 1301}, 125 (2013),
  arXiv:1210.8038.
\bibitem{Boito:2012cr} 
  D.~Boito, M.~Golterman, M.~Jamin, A.~Mahdavi, K.~Maltman, J.~Osborne and S.~Peris,
  Phys.\ Rev.\ D {\bf 85}, 093015 (2012),
  arXiv:1203.3146.


\end{thebibliography}
\end{document}